\documentclass[aps,onecolumn,showpacs,footinbib]{revtex4}%
\usepackage[latin9]{inputenc}
\usepackage{amssymb}
\usepackage{amsmath}
\usepackage{amscd}
\usepackage{graphicx}
\usepackage{subfigure}
\usepackage{float}
\begin{document}
\draft
\title{Simplified construction and physical realization of $n$-qubit controlled
phase gates}
\author{Shi-Biao Zheng\thanks{%
E-mail: sbzheng11@163.com}}
\address{Department of Physics\\
Fuzhou University\\
Fuzhou 350002, P. R. China}
\date{\today }

\begin{abstract}
We show that with the assistance of a third level of the qubits an n-qubit
phase gate can be constructed from $2n-4$ two-qutrit conditional swap gates,
a single qutrit-qubit controlled phase gate, and two single-qutrit
operations. Unlike previous schemes, our scheme uses the auxiliary level to
''expose'' some state to the qutrit-qubit controlled phase gate, instead of
using it to ''hide'' states from the conditional dynamics. Neither the
number of the additional levels nor that of single-qutrit operations needs
to increase with $n$. We propose a physical implementation of the required
elementary gates in cavity QED, and show that the total gate time may be
greatly reduced as compared with that required in the previous methods.
\end{abstract}


\vskip 0.5cm \maketitle \narrowtext

\section{INTRODUCTION}

A quantum computer, taking advantage of superposition and entanglement,
could realize additional information processing functions. Since Shor
discovered that a quantum computer could efficiently factorize large
integers in 1994 [1], quantum computation has become a truly
interdisciplinary field across physics, information science, and
engineering. In a quantum computer information is stored in quantum bits
(qubits) which are represented by two-level systems, such as atoms and ions.
The building blocks of a quantum computer are logic gates and any quantum
computational network can be decomposed into a series of two-qubit plus
one-qubit logic gates [2,3].

For the implementation of a practical quantum computational task, a large
number of qubits should be involved and controlled quantum gates among these
qubits are required. Of particular importance is the $n$-qubit controlled
phase gate that shifts the phase of one and only one of the state
components. This gate is an essential ingredient for implementation of
quantum algorithms [1,4] and quantum Fourier transform [5]. By placing the
Hadamard gates before and after the three-qubit controlled $\pi $-phase gate
on one of the qubits one can implement the Toffoli gate that inverts the
state of the target qubit conditional on the state of the two control
qubits. In addition, such phase gates are useful for the implementation of
quantum error correction [6-12]. The Toffoli gate has been demonstrated in
nuclear magnetic resonance [7], linear optics [13], ion trap [14], and
circuit QED systems [12,15], however the controlled phase gates involving
more than three qubits has not been experimentally implemented. Though an
n-qubit controlled phase gate could be decomposed into the elementary one-
and two-qubit gates, it would be extremely complex and difficult to solve a
practical problem, for example, a search with a quantum computer for an item
from a disordered system [4], via such a decomposition. On one hand, the
number of needed logic operations exponentially increases with the number of
qubits. On the other hand, a quantum system is very fragile and may be
destroyed by decoherence arising from the coupling with the environment. The
error of performance increases as the number of logic gates increases.

Recently, a highly efficient scheme has been proposed for implementation of
Grover's search algorithm in the trapped-ion system using Householder
reflections [16]. The distinct feature of this scheme is that each of the
inversion--about-average operation and the oracle query can be realized in a
single step so that the physical implementation of each logic iteration is
significantly simplified as compared with the methods based on decomposition
of multi-qubit controlled phase gates. However, the scheme requires the ions
to be initially prepared in the entangled Dicke state, which is
experimentally demanding. The search algorithm can also be simplified using
qudits [17]. Despite these advances, many theoretical and experimental
endeavors are still being directed toward the realization of multi-qubit
controlled phase gates for their applications in Shor's algorithm and
quantum error correction. So far implementation of the Toffoli gate and
three-qubit controlled phase gate based on one- and two-qubit gates has not
been reported due to decoherence, and it is of importance to simplify the
realization of a multi-qubit gate so that the number of required elementary
operations does not exponentially increase with the number of qubits. Resch
et al. have shown that the number of two-qubit gates required to implement a
Toffoli gate acting on three qubits can be reduced if one of the three
qubits has a third state that is accessible during the gate operation [18].
The basic idea of the method is to ''hide'' certain states from the
two-qubit controlled phase gate. The technique can be generalized to
higher-order Toffoli gate with $n$ qubits by making the target qubit an $n$%
-level qudit. In general, this method requires one two-qubit controlled
phase gate, $2n-4$ controlled-NOT gates, and $2(n-2)$ single-qudit gates to
construct an $n$-qubit controlled phase gate. The limitation of the method
is that it requires the number of the accessible states of the target to
equal the number of the qubits involved in the gate operation, which is
experimentally problematic since it may be difficult to get as many states
with long coherence times as required in an realistic physical system.

In this paper we show that, with the assistance of an auxiliary state, an
n-qubit quantum phase gate could be constructed from $2n-4$ two-qutrit
conditional swap gates, a single qutrit-qubit controlled phase gate, and two
single-qutrit gates. Unlike the previous methods [12-15,18,19], the
auxiliary state is used to ''expose'' some state to the two-qutrit
controlled phase gate, not to ''hide'' certain states. In comparison with
the method of Ref. [18], the present one does not require the number of the
available states of the target increase with $n$. For each qubit only one
additional state is required to be accessible during the gate operation.
Furthermore, for certain realistic physical systems it is easier to
implement the conditional swap gate than the controlled-NOT gate. Another
advantage of the present method is that the number of required single-qutrit
operations is independent of $n$. We propose an experimental realization of
the two-qutrit conditional swap gate and qutrit-qubit controlled phase gate
in cavity QED. During the operation the atomic qubits are always in the
ground states and the cavity mode is only virtually excited and thus the
scheme is insensitive to both the atomic spontaneous emission and cavity
decay. The scheme is generic and can be implemented in other physical
systems in which the qubits have an auxiliary state.

The paper is organized as follows. In Sec.2, we describe the method to
construct the $n$-qubit controlled phase gate using only qutrits, and show
that the number of required single-qutrit gates does not increase with $n$.
In Sec.3, as an example for the physical implementation of this method we
demonstrate that the required elementary gates can be realized in the
context of cavity QED. The cavity mode, together with external classical
fields, can induce the controlled atom-atom coupling. It is shown that in
this system the implementation of the conditional swap gate is easier than
that of the controlled-NOT\ gate and the gate time is reduced. The
conclusion appears in Sec.4.

\section{CONSTRUCTION OF THE n-QUBIT CONTROLLED PHASE GATE WITH QUTRITS}

We first consider a three-qubit system. The computational basis states of
each qubit is represented by $\left| 1\right\rangle $ and $\left|
0\right\rangle $. Meanwhile, each qubit has an auxiliary state $\left|
a\right\rangle $. The main ingredients for constructing the three-qubit
controlled phase gate are the two-qutrit conditional swap gate $%
U_{j,k}=e^{\pi (\left| 1_ja_k\right\rangle \left\langle a_j1_k\right|
-\left| a_j1_k\right\rangle \left\langle 1_ja_k\right| )/2}$ and
qutrit-qubit controlled phase gate $V_{j,k}=e^{i\phi \left|
a_j1_k\right\rangle \left\langle a_j1_k\right| }$. Without loss of the
generality, we assume that the three qubits are initially in the state

\begin{equation}
\sum_{x,y,z=0,1}\alpha _{x,y,z}\left| x_1y_2z_3\right\rangle .
\end{equation}
We first perform the single-qutrit transformation $L_1$ on qutrit 1$:\left|
1_1\right\rangle \rightarrow \left| a_1\right\rangle $, which leads to
\begin{equation}
\sum_{y,z=0,1}\alpha _{0,y,z}\left| 0_1y_2z_3\right\rangle
+\sum_{y,z=0,1}\alpha _{1,y,z}\left| a_1y_2z_3\right\rangle .
\end{equation}
Then the gate $U_{12}$ is performed between qutrits 1 and 2, resulting in
\begin{equation}
\sum_{y,z=0,1}\alpha _{0,y,z}\left| 0_1y_2z_3\right\rangle
+\sum_{z=0,1}\alpha _{1,0,z}\left| a_10_2z_3\right\rangle
+\sum_{z=0,1}\alpha _{1,1,z}\left| 1_1a_2z_3\right\rangle .
\end{equation}
Next we apply the gate $V_{23}$ between 2 and 3 to obtain
\begin{equation}
\sum_{y,z=0,1}\alpha _{0,y,z}\left| 0_1y_2z_3\right\rangle
+\sum_{z=0,1}\alpha _{1,0,z}\left| a_10_2z_3\right\rangle +\alpha
_{1,1,0}\left| 1_1a_20_3\right\rangle +\alpha _{1,1,1}e^{i\phi }\left|
1_1a_21_3\right\rangle .
\end{equation}
Now the gate $U_{21}$ is again performed between 1 and 2, leading to
\begin{equation}
\sum_{y,z=0,1}\alpha _{0,y,z}\left| 0_1y_2z_3\right\rangle
+\sum_{z=0,1}\alpha _{1,0,z}\left| a_10_2z_3\right\rangle +\alpha
_{1,1,0}\left| a_11_20_3\right\rangle +\alpha _{1,1,1}e^{i\phi }\left|
a_11_21_3\right\rangle .
\end{equation}
Finally, we perform the single-qutrit transformation $M_1:$ $\left|
a_1\right\rangle \rightarrow \left| 1_1\right\rangle $ and the state becomes
\begin{equation}
\sum_{y,z=0,1}\alpha _{0,y,z}\left| 0_1y_2z_3\right\rangle
+\sum_{z=0,1}\alpha _{1,0,z}\left| 1_10_2z_3\right\rangle +\alpha
_{1,1,0}\left| 1_11_20_3\right\rangle +\alpha _{1,1,1}e^{i\phi }\left|
1_11_21_3\right\rangle ,
\end{equation}
in which if and only if all the qubits are initially in the state $\left|
1\right\rangle $ the system undergoes a phase shift $\phi $. It is
worthwhile mentioning that only when qubits 1 and 2 are initially in the
state $\left| 1_11_2\right\rangle $ the state of qubit 2 can be transformed
to $\left| a_2\right\rangle $ by the gate $U_{12}$ and then qubits 2 and 3
be subjected to the controlled phase gate $V_{23}$ which only affects the
non-computational state $\left| a_21_3\right\rangle $. In other words, the
auxiliary level $\left| a\right\rangle $ is used to ''expose'' the initial
qubit state $\left| 1_11_21_3\right\rangle $ to the controlled phase gate.
This is distinguished from the previous schemes [12-15,18,19] in which the
auxiliary levels are used to ''hide'' certain states so that the controlled
phase gate only affects one computational state.

We note that the idea can be generalized to produce the $n$-qubit phase gate
\begin{equation}
U_n=e^{i\phi \left| 1_11_2...1_n\right\rangle \left\langle
1_11_2...1_n\right| }
\end{equation}
by applying a sequence of operations: $%
L_1,U_{1,2},U_{2,3},...,U_{n-2,n-1},V_{n-1,n},U_{n-1,n-2},U_{n-2,n-3},...U_{2,1},
$ and $M_1$. Therefore, $2n-4$ two-qutrit conditional swap operations, a
single qutrit-qubit controlled phase gate, and two single-qutrit operations
are sufficient for the construction of an $n$-qubit controlled phase gate.
One appealing feature of the method is that neither the number of the
required additional levels nor the number of single-qutrit operations
increases with $n$.

\section{PHYSICAL IMPLEMENTATION}

We consider that $n$ identical atoms are trapped in a cavity. Each atom has
one excited state $\left| r\right\rangle $ and three ground states $\left|
1\right\rangle $, $\left| 0\right\rangle $, and $\left| a\right\rangle $, as
shown in Fig. 1. The transition $\left| 1\right\rangle \rightarrow \left|
r\right\rangle $ is coupled to the cavity mode with the coupling constant $g$%
. For implementation of logic operations between the jth and (j+1)th atoms,
the transition $\left| a\right\rangle \rightarrow \left| r\right\rangle $
for each of these two atoms is driven by a classical laser field. Assume the
classical field and cavity mode are detuned from the respective transitions
by $\Delta _1$ and $\Delta _2$, respectively. In the interaction picture,
the Hamiltonian is
\begin{equation}
H=e^{i\Delta _1t}(\Omega _je^{-i\varphi _j}\left| r_j\right\rangle
\left\langle a_j\right| +\Omega _{j+1}e^{-i\varphi _{j+1}}\left|
r_{j+1}\right\rangle \left\langle a_{j+1}\right| )+\sum_{m=1}^ngae^{i\Delta
_2t}\left| r_m\right\rangle \left\langle 1_m\right| +H.c.,
\end{equation}
where $a$ is the annihilation operator of the cavity mode, and $\Omega _j$
and $\varphi _j$ are the Rabi frequency and phase of the laser field driving
the $j$th atom, respectively. Under the condition $\Delta _1$, $\Delta _2\gg
\Omega _j,g$ the upper level $\left| r\right\rangle $ can be adiabatically
eliminated, leading to the Raman coupling of the two ground states and Stark
shifts. Then the dynamics of the system is described by the effective
Hamiltonian [20]
\begin{eqnarray}
H_e &=&-\frac{\Omega _j^2}{\Delta _1}\left| a_j\right\rangle \left\langle
a_j\right| -\frac{\Omega _{j+1}^2}{\Delta _1}\left| a_{j+1}\right\rangle
\left\langle a_{j+1}\right| -\lambda _j(aS_j^{+}e^{i\varphi _j}e^{i\delta
t}+a^{\dagger }S_j^{-}e^{-i\varphi _j}e^{-i\delta t})  \nonumber \\
&&-\lambda _{j+1}(aS_{j+1}^{+}e^{i\varphi _{j+1}}e^{i\delta t}+a^{\dagger
}S_{j+1}^{-}e^{-i\varphi _{j+1}}e^{-i\delta t})-\sum_{m=1}^n\frac{g^2}{%
\Delta _2}a^{+}a\left| 1_m\right\rangle \left\langle 1_m\right| ,
\end{eqnarray}
where $\lambda _j=\frac{\Omega _jg}2(\frac 1{\Delta _1}+\frac 1{\Delta _2})$%
, $\delta =\Delta _2-\Delta _1,$ $S_j^{+}=\left| a_j\right\rangle
\left\langle 1_j\right| $, and $S_j^{-}=\left| 1_j\right\rangle \left\langle
a_j\right| .$

In the case $\delta \gg \lambda _j$, $\frac{\Omega _j^2}{\Delta _1}$, $\frac{%
g^2}{\Delta _2}$, there is no energy exchange between the atomic system and
the cavity. The energy conserved transitions are between $\left|
a_j1_{j+1}n\right\rangle $ and $\left| 1_ja_{j+1}n\right\rangle $. The
effective coupling for the transition $\left| 1_ja_{j+1}n\right\rangle
\rightarrow \left| a_j1_{j+1}n\right\rangle $, mediated by $\left|
1_j1_{j+1}n+1\right\rangle $ and $\left| a_ja_{j+1}n-1\right\rangle $ is
given by [21,22]
\begin{eqnarray}
&&\frac{\left\langle a_j1_{j+1}n\right| H_e\left| 1_j1_{j+1}n+1\right\rangle
\left\langle 1_j1_{j+1}n+1\right| H_e\left| 1_ja_{j+1}n\right\rangle }\delta
\nonumber  \label{4} \\
&&\ \ +\frac{\left\langle a_j1_{j+1}n\right| H_e\left|
a_ja_{j+1}n-1\right\rangle \left\langle a_ja_{j+1}n-1\right| H_e\left|
1_ja_{j+1}n\right\rangle }{-\delta }  \nonumber \\
\ &=&\xi e^{i\varphi },
\end{eqnarray}
where $\xi =\frac{\lambda _j\lambda _k}\delta $ and $\varphi =\varphi
_j-\varphi _{j+1}$. Since the two transition paths interfere destructively
the effective Rabi frequency is independent of the photon-number of the
cavity mode. In addition to the two-qubit coupling, the nonresonant Raman
coupling leads to further Stark shift. Then we obtain the new effective
Hamiltonian
\begin{eqnarray}
H_e^{^{\prime }} &=&(-\frac{\Omega _j^2}{\Delta _1}+\frac{\lambda _j^2}\delta
aa^{\dagger })\left| a_j\right\rangle \left\langle a_j\right| +(-\frac{%
\Omega _{j+1}^2}{\Delta _1}+\frac{\lambda _{j+1}^2}\delta aa^{\dagger
})\left| a_{j+1}\right\rangle \left\langle a_{j+1}\right|  \nonumber \\
&&-\frac{\lambda _j^2}\delta a^{\dagger }a\left| 1_j\right\rangle
\left\langle 1_j\right| -\frac{\lambda _{j+1}^2}\delta a^{\dagger }a\left|
1_{j+1}\right\rangle \left\langle 1_{j+1}\right|  \nonumber \\
&&+\xi (e^{i\varphi }S_j^{+}S_{j+1}^{-}+e^{-i\varphi
}S_j^{-}S_{j+1}^{+})-\sum_{m=1}^n\frac{g^2}{\Delta _2}a^{+}a\left|
1_m\right\rangle \left\langle 1_m\right| .
\end{eqnarray}
The photon-number does not change during the process since $[a^{\dagger
}a,H_e^{^{\prime }}]=0$ . When the cavity mode is initially in the vacuum
state it will remains in the vacuum state throughout the procedure. Then the
effective Hamiltonian $H_e^{^{\prime }}$ reduces to
\begin{equation}
H_e^{^{\prime }}=-\mu _j\left| a_j\right\rangle \left\langle a_j\right| -\mu
_{j+1}\left| a_{j+1}\right\rangle \left\langle a_{j+1}\right| +\xi
(e^{i\varphi }S_j^{+}S_{j+1}^{-}+e^{-i\varphi }S_j^{-}S_{j+1}^{+}),\text{ }
\end{equation}
where

\[
\mu _j=\frac{\Omega _j^2}{\Delta _1}-\frac{\lambda _j^2}\delta
\]
After an interaction time t we obtain the state evolution
\begin{eqnarray}
\left| a_j1_{j+1}\right\rangle &\rightarrow &e^{i\mu t}\{[\cos (\eta t)-i%
\frac \epsilon {2\eta }\sin (\eta t)]\left| a_j1_{j+1}\right\rangle -i\frac
\xi \eta e^{-i\varphi }\sin (\eta t)\left| 1_ja_{j+1}\right\rangle \},
\nonumber \\
\left| 1_ja_{j+1}\right\rangle &\rightarrow &e^{i\mu t}\{[\cos (\eta t)+i%
\frac \epsilon {2\eta }\sin (\eta t)]\left| 1_ja_{j+1}\right\rangle -i\frac
\xi \eta e^{i\varphi }\sin (\eta t)\left| a_j1_{j+1}\right\rangle \},
\nonumber \\
\left| a_j0_{j+1}\right\rangle &\rightarrow &e^{i\mu _jt}\left|
a_j0_{j+1}\right\rangle .
\end{eqnarray}
where $\mu =(\mu _j+\mu _{j+1})/2$, $\epsilon =\mu _{j+1}-\mu _j$ and $\eta =%
\sqrt{\xi ^2+(\mu _j-\mu _{j+1})^2/4}$. The basis states $\left|
0_j0_{j+1}\right\rangle $, $\left| 0_j1_{j+1}\right\rangle $, $\left|
1_j0_{j+1}\right\rangle $, and $\left| 1_j1_{j+1}\right\rangle $ remain
unchanged during the interaction. As has been shown, before the operation $%
U_{j,j+1}$ ($U_{j+1,j}$) the jth and (j+1)th atoms have no probability of
being populated in the states $\left| 0_ja_{j+1}\right\rangle ,\left|
1_ja_{j+1}\right\rangle $ ($\left| a_j1_{j+1}\right\rangle $), and $\left|
a_ja_{j+1}\right\rangle $ so that it is unnecessary to consider the
evolution of these states during the gate operation $U_{j,j+1}$ ($U_{j+1,j}$%
). With the choice of $\mu _j=\mu _{j+1}=\mu ,$ $\xi t=\pi /2$ and $\varphi
=-\pi /2$ ($\varphi =\pi /2$) the conditional swap operation $U_{j,j+1}$ ($%
U_{j+1,j}$) is obtained through the transformation (13) plus the
single-qubit phase shifts: $\left| a_j\right\rangle \longrightarrow e^{-i\pi
\mu /2\xi }\left| a_j\right\rangle $ and $\left| a_{j+1}\right\rangle
\longrightarrow e^{-i\pi \mu /2\xi }\left| a_{j+1}\right\rangle $. Before
the operation $V_{n-1,n}$ the $(n-1)$th and $n$th atoms are not populated in
the states $\left| 0_{n-1}a_n\right\rangle ,\left| 1_{n-1}a_n\right\rangle $%
, and $\left| a_{n-1}a_n\right\rangle $. Therefore, we only need to consider
the evolutions of $\left| a_{n-1}1_n\right\rangle $ and $\left|
a_{n-1}0_n\right\rangle $. With the choice of $\eta t=\pi $, we obtain
\begin{eqnarray}
\left| a_{n-1}1_n\right\rangle &\rightarrow &e^{i\pi (1+\mu /\eta )}\left|
a_{n-1}1_n\right\rangle ,  \nonumber \\
\left| a_{n-1}0_n\right\rangle &\rightarrow &e^{i\pi \mu _{n-1}/\eta }\left|
a_{n-1}1_n\right\rangle .
\end{eqnarray}
This transformation, together with the single-qutrit phase shifts $\left|
a_{n-1}\right\rangle \longrightarrow e^{-i\pi \mu _{n-1}/\eta }\left|
a_{n-1}\right\rangle $, corresponds to the conditional phase operation $%
V_{n-1,n}$. The conditional phase shift $\pi (1+\epsilon /2\eta )$ is
controllable via the Rabi frequencies of the two classical fields.

It is worth noticing that the qutrit-qubit controlled phase gate reduces to
the two-qubit controlled phase gate when the logic states of qubit $n-1$ are
represented by $\left| 0_{n-1}\right\rangle $ and $\left|
a_{n-1}\right\rangle $ and qubit $n$ uses $\left| 0_n\right\rangle $ and $%
\left| 1_n\right\rangle $ as the logic states [21,22]. For the present
cavity QED system the duration of the two-qutrit conditional swap gate is
one half of of that of the two-qubit controlled $\pi -$phase gate. For the
implementation of the $n$-qubit controlled $\pi -$phase gate in the present
system the scheme of Ref. [18] requires $2n-3$ two-qubit controlled $\pi -$%
phase gates, $2n-4$ single-qubit gate, and $2(n-2)$ single-qudit gates. The
present method instead uses $2n-4$ two-qutrit conditional swap gates, one
controlled qutrit-qubit controlled $\pi -$phase gate, and two single-qutrit
gates. The total time for two-atom couplings is reduced by $(2n-4)\pi /(2\xi
)$. In addition, the number of required single-atom operations is greatly
reduced. The method of Ref. [15] uses a single two-qubit controlled phase
gate and two qubit-qutrit swap gate to implement the three-qubit phase gate.
However, the scheme can not be directly generalized to higher-order $n$%
-qubit controlled phase gates. Furthermore, the duration of the second swap
gate should be three times of that of the first one.

\section{SUMMARY}

In summary, we have suggested a scheme for decomposing an $n$-qubit
controlled phase gate into $2n-4$ two-qutrit conditional swap gates, a
single qutrit-qubit controlled phase gate, and two single-qutrit gates. In
the scheme each qubit needs to have a single additional state that can be
addressed during the gate operation. This auxiliary state is used to
''expose'' one and only one initial computational state to the qutrit-qubit
controlled phase gate, which is distinguished from previous methods using
the additional levels to ''hide'' one or more computational states from the
two-qubit controlled phase gate. In comparison with the scheme of Ref. [18],
the procedure is greatly simplified and the total gate time is reduced. We
illustrate the idea in cavity QED. However, the scheme can be readily
applied to other systems that have three levels with long coherence times,
such as trapped ions and superconducting circuits.

This work was supported by the Major State Basic Research Development
Program of China under Grant No. 2012CB921601, National Natural Science
Foundation of China under Grant No. 10974028, the Doctoral Foundation of the
Ministry of Education of China under Grant No. 20093514110009, and the
Natural Science Foundation of Fujian Province under Grant No. 2009J06002.

Fig. 1 (color online). The atomic level configuration and excitation scheme
to realize the two-qutrit swap gate and qutrit-qubit controlled phase gate.
The transition $\left| 1\right\rangle \rightarrow \left| r\right\rangle $ is
coupled to the cavity mode and $\left| a\right\rangle \rightarrow \left|
r\right\rangle $ is driven by a classical laser field. The classical field
and cavity mode are detuned from the respective transitions by $\Delta _1$
and $\Delta _2$.
\begin{figure}[C]
\includegraphics[width=0.5\columnwidth]{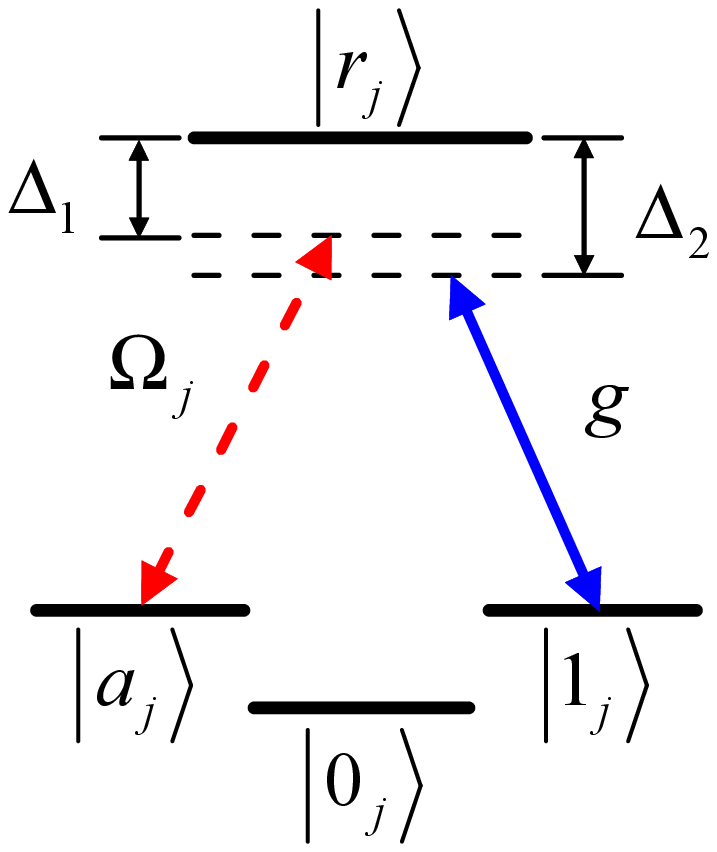}
\caption{}
\end{figure}
\end{document}